\documentclass{aa}

\usepackage{graphics}

\begin{document}
  
\title{Long-term photometry of the Wolf-Rayet stars 
WR 137, WR 140, WR 148, and WR 153
\thanks{Based on observations collected at the National
Astronomical Observatory Rozhen, Bulgaria}}

\author{Kiril P. Panov\inst{1},
Martin Altmann\inst{2},
\and Wilhelm Seggewiss\inst{2}}

\institute{Institute of Astronomy, Bulgarian Academy 
of Sciences, Sofia, Bulgaria \and
Universit\"atssternwarte Bonn, Auf dem H\"ugel 71, D-53121 Bonn, Germany}

\date{Received 31 August 1999/18 Januar 2000}

\thesaurus{03.20.4:08.02.2;08.02.4;08.03.4;08.23.2}

\offprints{kpanov@astro.bas.bg}

\maketitle

\markboth{K.~P. Panov et~al., Long-term photometry of WR stars}
      	{K.~P. Panov et~al., Long-term photometry of WR stars}

\begin{abstract}
In 1991, a long term $UBV$-photometry campaign of four Wolf-Rayet stars
was started using the 60 cm telescope of the National Astronomical 
Observatory Rozhen, Bulgaria.
Here we report on our observational results and discuss the
light variations.

The star WR\,137 was observed during 1991 - 1998. 
No indications of eclipses were found, though random 
light variations with small amplitudes
exist, which are probably due to dynamical wind instabilities. 

WR\,140 was also monitored between 1991 and 1998. 
In 1993, a dip in the light curve in all passbands
was observed shortly after periastron passage, 
with amplitude of 0.03\,mag in $V$. This is interpreted 
in terms of an ``eclipse'' by dust condensation
in the WR-wind. The amplitude  
of the eclipse increases towards shorter 
wavelengths; thus, electron scattering alone is 
not sufficient to explain the observations. An additional   
source of opacity is required, possibly  Rayleigh 
scattering. After the eclipse, the light in all 
passbands gradually increased to reach the 
``pre-eclipse'' level in 1998.
The very broad shape of the light minimum suggests 
that a dust envelope was built up around  the WR-star 
at periastron passage by wind-wind interaction,
and was gradually dispersed after 1993.   

Our observations of WR\,148 (WR + c?) confirm 
the 4.3 d period; however, they also show 
additional significant scatter.
Another interesting finding is a long-term variation 
of the mean light (and, possibly, of the amplitude) on a time scale
of years. There is some indication of a 4 year cycle of
that long-term variation. We discuss the 
implications for the binary model. 

Our photometry of WR\,153  is consistent 
with the quadruple model of this star 
by showing that both orbital periods, 
6.7 d (pair A) and 3.5 d (pair B),
exist in the light variations. A search in the HIPPARCOS   
photometric data also reveals both
periods, which is an independant confirmation. 
No other periods in the light variability of that  
star are found. The longer period light curve shows     
only one minimum, which might be due to an atmospheric eclipse;  
the shorter period light curve shows two minima, 
indicating that both stars in pair B are eclipsing each other.                                                                                   
\end{abstract}

\section{Introduction}

\begin{table*}
\caption[]{Summary of data for the program stars. The last
two columns contain\\ the emission line-contribution to the 
colours (Pyper 1966)}
\label{summary}
\begin{tabular}{rrcrrrr}
\hline\noalign{\smallskip}
  WR & HD & Spectral & Comparison & Check & $\delta(B-V)$ & $\delta(U-B)$ \\ 
     &    & types    & star, HD   & star, HD & [mag] & [mag] \\
  \noalign{\smallskip}
\hline
  \noalign{\smallskip}
137  & 192641 & WC7+abs  &  192538 & 192987 & +0.07 & $-0.11$ \\
140  & 193793 & WC7+O4-5 &  193888 & 193926 & +0.05 & $-0.12$ \\
148  & 197406 & WN7+c?   &  197619 & 196939 & +0.01 & $-0.01$ \\
153  & 211853 & 2$\times$WN+O, or & 211430 & $---$ & +0.02 & $-0.02$ \\
     &      & WN+O and O+O &        &       &       &         \\
  \noalign{\smallskip}
\hline
\end{tabular}
\end{table*}

Photometric studies of Wolf-Rayet (WR) stars 
during the past decades (e.g. Moffat \& Shara 1986; Lamontagne 
\& Moffat 1987; van Genderen et al. 1987; Balona 
et al. 1989; Robert et al. 1989; Gosset et al. 1990;
Antokhin et al. 1995; Marchenko et al. 1998a;
Marchenko et al. 1998b) have revealed 
light variations of several per cent 
(up to 0.1 mag) on time-scales (typically) 
of days. WR stars are generally 
believed to be evolved Population I stars, 
descendants of Of-type stars (Maeder 1996). 
They exhibit strong, dense winds (mass loss rates 
of $10^{-5}$  to  $10^{-4}$ M$_{\sun}$ yr$^{-1}$) which, in most cases,
hide the stellar surface. The wind-flow is 
dependant on time. Moffat et al. (1988,
1994) and Robert (1994) discovered the existence 
of small, outward moving wind condensations, which they called
propagating blobs. Unlike most O-type stars, the continuum 
light of many WR stars originates from a layer
in the dense wind ($\tau = 1$), a ``pseudo-photosphere'' 
(van Genderen et al. 1987), which could be 
inhomogeneous because of dynamical wind instabilities. 
The brightness variations of some WR stars proved to be periodic 
and are possibly due to binary or rotation 
effects. Core (photospheric) eclipses as well as
atmospheric eclipses have been observed. The 
latter are characterized by only one V-shaped minimum on 
the light curve, which is caused by the atmospheric 
eclipse of an O-type star by the
WR star's extended wind (Lamontagne et al. 1996). 
Random light variations are common in WR stars 
and they are often superimposed on the regular 
(binary) variations, increasing the ``noise'' and
sometimes even totally disturbing the underlying 
regular light variations. 
Marchenko et al. (1998b) suggested that random light
variations (light scatter) may be caused 
by short-lived, core-induced, multimode fluctuations,
propagating in the wind. Other causes of variability, 
such as radial pulsations (Maeder 1985)
non-radial pulsations (Vreux 1985, Antokhin et al. 
1995, Rauw et al. 1996) and axial rotation 
(Matthews \& Moffat 1994) have been proposed for WR stars. 
Occasional ``eclipses'' caused by dust formation in 
late-type WC stars have been studied by Veen et al. (1998).

WR\,137 is a well known dust maker (Williams 1997; 
Marchenko et al. 1999). However little is known 
about long-term light variations for that star and 
its binary status is still uncertain.
WR\,140 is another repeating dust maker 
(Williams 1997). The orbit is well determined.
Because of its high eccentricity ($e = 0.84$)  the
strongest wind interaction occurs at periastron
passage. During the last periastron passage in 1993,
WR\,140 received much attention and has been studied at
different wavelengths from X-ray to radio. 
However, only a few photometric studies in the
optical were carried out so far and the long-term 
behaviour of that star is not known. 
Both stars WR\,137 and WR\,140 are included in the
infrared study by Williams et al. (1987a) and
reported to have dust shells.

WR\,148 is a good candidate for a  
WR + c (WR plus compact companion) binary. There is some  
controversy about the light variations concerning the period
and the shape of the light curve. Marchenko et al. 
(1998a) were not able to detect the 4.31 d
binary period in the HIPPARCOS photometry data, 
otherwise well known from ground-based      
observations (Marchenko et al. 1996). The very
``noisy'' light curve and unusually broad 
minimum need further investigation.

WR\,153 is a quadruple system (Massey 1981), containing 
a WN + O and an O + O system, or
two WN + O pairs (Panov \& Seggewiss
1990). During the past  years, several photometric 
studies have been carried out. Yet the light  
variability of the two pairs could not always be 
unambiguously separated (Lamontagne et al. 1996).
Our aim is to try to solve some of these controversial questions.

\section{Observations}

In 1991 we started a long-term photometric study of the
four WR stars at the National Astronomical Observatory Rozhen,
Bulgaria, using the 60 cm telescope and the UBV single channel,
photon counting photometer.
The photometric equipment has been used
for many years and proved to be very
stable (cf. Panov et al. 1982).

Table~1 contains the comparison and the check
stars used. Generally, a 20\arcsec\,diaphragm and an integration
of 10\,s were used. Each measurement consists of four consecutive 
integration cycles. An observing cycle was arranged in the 
following way: Sky - Comp - WR - Comp - Sky
and was repeated 3 to 5 times,
depending on the quality of the night. A separate
measurement of the comparison star against
the check star in the same way was obtained
before or after the WR star observation.
Thus a nightly mean was
calculated from the 3 to 5 individual measurements.
The standard error of the nightly mean is 
0.003 - 0.005 mag in most cases.
Reduction of the data was made taking into account
dead-time effects, atmospheric extinction, and transformation
into the standard $UBV$ system. In the following tables
we present the magnitudes in the standard system; for WR\,137,
WR\,140, and WR\,148 the data are magnitude
differences in the sense: comparison
star minus WR star. The contribution $\delta(B-V)$ and $\delta(U-B)$
of emission lines to the respective colours are taken from
Pyper (1966) and included in Table~1 (last
two columns). No corrections have been applied
for emission lines in our data. However, it does
seem possible to distinguish the continuum
light from the emission line variations by
comparing the light in the $UBV$ passbands.
              
Many WR stars show subtle short and long term variations 
as can be seen in the case of
WR\,148 (Sec. 3.3). Therefore it is preferable
to use the same photometric equipment for long term studies. This reduces
possible systematic effects caused by slightly different passbands or
response curves.

\begin{figure}
  \centering
  \resizebox{08cm}{!}{\includegraphics{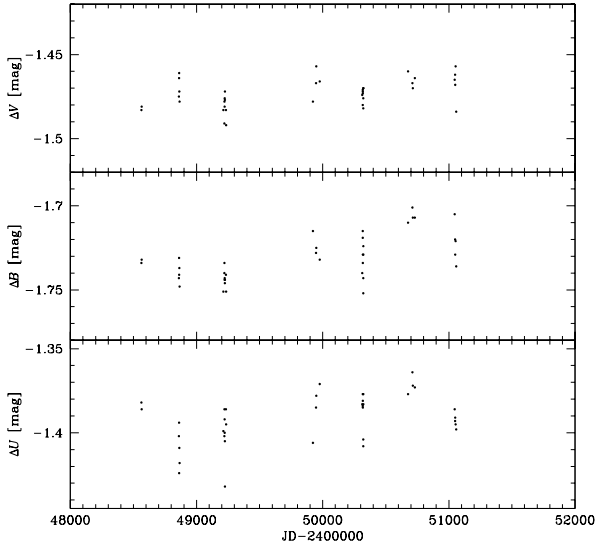}}
  \hfill
    \caption{Light curves of WR\,137 (data from Table 2)}
\end{figure}

\begin{figure}
  \centering
  \resizebox{8cm}{!}{\includegraphics{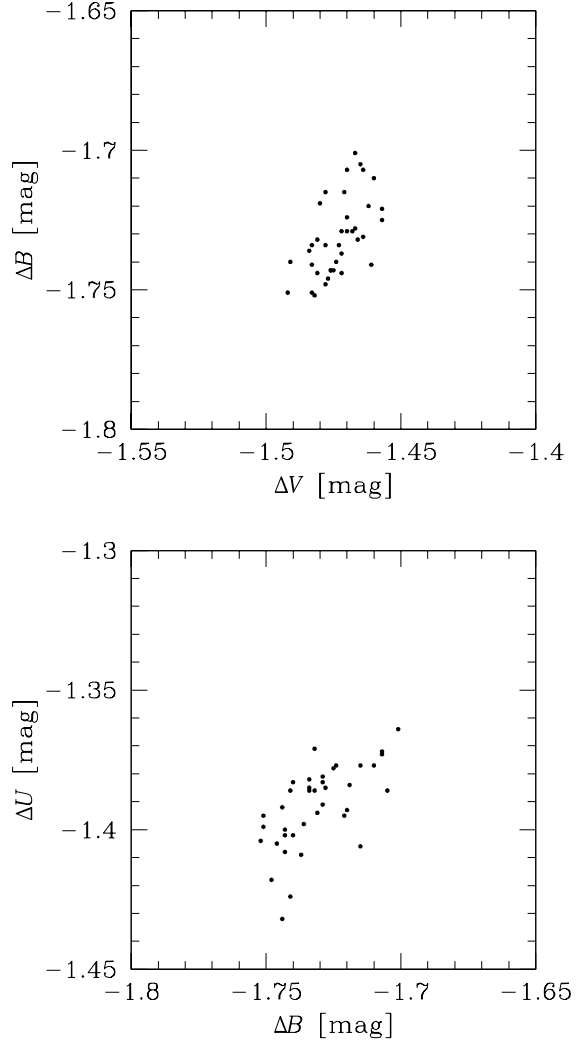}}
  \hfill
    \caption{Random light variability correlations for WR\,137
     (data from Table 2).}
\end{figure}

\begin{table}
\caption[]{Differential photometry of WR\,137
(= HD 192641)  --  in the sense comparison
star HD 192538 minus WR\,137}
   \setlength\tabcolsep{10pt}
\renewcommand{\baselinestretch}{0.8}
\small
\begin{tabular}{rcccc}
\hline
\noalign{\smallskip}
Year  &  JD-2400000 &    $\Delta V$   &     $\Delta B$    &    $\Delta U$ \\
      &             &   [mag]  &    [mag]   &  [mag]  \\
\noalign{\smallskip}
\hline
\noalign{\smallskip}
1991  & 48563.225   & $-$1.483   &  $-$1.734    & $-$1.382 \\
      & 48565.229   & $-$1.481   &  $-$1.732    & $-$1.386 \\
\noalign{\smallskip}
\hline
\noalign{\smallskip}
1992  & 48860.395   &  $-$1.475  &  $-$1.743    & $-$1.402 \\
      & 48861.395   &  $-$1.464  &  $-$1.731    & $-$1.394 \\
      & 48862.429   &  $-$1.461  &  $-$1.741    & $-$1.424 \\
      & 48863.410   &  $-$1.472  &  $-$1.737    & $-$1.409 \\
      & 48865.367   &  $-$1.478  &  $-$1.748    & $-$1.418 \\
\noalign{\smallskip}
\hline
\noalign{\smallskip}
1993  & 49212.437   &  $-$1.483  &  $-$1.751    & $-$1.399 \\
      & 49220.396   &  $-$1.491  &  $-$1.740    & $-$1.402 \\
      & 49221.359   &  $-$1.478  &  $-$1.734    & $-$1.386 \\
      & 49222.392   &  $-$1.481  &  $-$1.744    & $-$1.392 \\
      & 49223.375   &  $-$1.476  &  $-$1.743    & $-$1.400 \\
      & 49224.392   &  $-$1.477  &  $-$1.746    & $-$1.405 \\
      & 49225.398   &  $-$1.472  &  $-$1.744    & $-$1.432 \\
      & 49233.341   &  $-$1.483  &  $-$1.741    & $-$1.386 \\
      & 49234.354   &  $-$1.492  &  $-$1.751    & $-$1.395 \\
\noalign{\smallskip}
\hline
\noalign{\smallskip}
1995  & 49922.475   &  $-$1.478  &  $-$1.715    & $-$1.406 \\
      & 49947.449   &  $-$1.467  &  $-$1.728    & $-$1.385 \\
      & 49949.428   &  $-$1.457  &  $-$1.725    & $-$1.378 \\
      & 49976.379   &  $-$1.466  &  $-$1.732    & $-$1.371 \\
\noalign{\smallskip}
\hline
\noalign{\smallskip}
1996  & 50313.390   &  $-$1.474  &  $-$1.740    & $-$1.383 \\
      & 50317.344   &  $-$1.480  &  $-$1.719    & $-$1.384 \\
      & 50317.361   &  $-$1.471  &  $-$1.715    & $-$1.377 \\
      & 50318.343   &  $-$1.472  &  $-$1.729    & $-$1.381 \\
      & 50318.363   &  $-$1.473  &  $-$1.734    & $-$1.385 \\
      & 50321.358   &  $-$1.482  &  $-$1.752    & $-$1.404 \\
      & 50321.375   &  $-$1.476  &  $-$1.743    & $-$1.408 \\
      & 50322.316   &  $-$1.470  &  $-$1.724    & $-$1.377 \\
      & 50322.338   &  $-$1.470  &  $-$1.729    & $-$1.383 \\
\noalign{\smallskip}
\hline
\noalign{\smallskip}
1997  & 50676.374   &  $-$1.460  &  $-$1.710    & $-$1.377 \\
      & 50711.311   &  $-$1.467  &  $-$1.701    & $-$1.364 \\
      & 50714.272   &  $-$1.470  &  $-$1.707    & $-$1.372 \\
      & 50730.268   &  $-$1.464  &  $-$1.707    & $-$1.373 \\
\noalign{\smallskip}
\hline
\noalign{\smallskip}
1998  & 51046.379   &  $-$1.465  &  $-$1.705    & $-$1.386 \\
      & 51049.343   &  $-$1.462  &  $-$1.720    & $-$1.393 \\
      & 51050.331   &  $-$1.468  &  $-$1.729    & $-$1.391 \\
      & 51053.334   &  $-$1.457  &  $-$1.721    & $-$1.395 \\
      & 51054.310   &  $-$1.464  &  $-$1.726    & $-$1.386 \\
      & 51058.364   &  $-$1.484  &  $-$1.736    & $-$1.398 \\
\noalign{\smallskip}
\hline
\end{tabular}
\end{table}

\section{Discussion}

\subsection{WR\,137}

WR\,137 = HD 192641 (WC7 + ?) has been studied in the 
infrared (IR) and peaks in brightness were 
reported in 1984.5 and in 1997, 
probably caused by heated dust (Williams 1997). 
The dust emission has been directly IR-imaged at two epochs
recently using the Hubble Space Telescope by
Marchenko et al. (1999).
The repetition of IR maxima occurs with a $\sim$13 yr
period, suggesting a possible binary origin, as found
for other WR periodic dust makers. WR\,137 was 
discovered to be a spectroscopic binary by Annuk (1995). 
However, Underhill (1992) did not find any evidence for 
binary motion in her data. Therefore, the binary status of
WR\,137 remains uncertain. 

Marchenko \& Pikhun (1992)
published a long-term photometric study for 1958 - 1989,
but it is based on photographic plates and the 
accuracy is insufficient to reveal light variations
below a few per cent. Our photometry is presented
in Table 2 and the light curves are shown in Fig 1.     
We searched for periodicities using the procedure of Lafler 
\& Kinman (1965), in the period range from 1 d to 100 d,
 but no period could be found.
The only photometric variations we can see in
our data are random light variations with amplitudes
of 0.02 mag (peak to peak) in $V$ during each observing 
season and up to 
0.03 mag (peak to peak) when we compare different years. 
(However, the peak to peak amplitude from all data is
0.05\,mag in $B$, and 0.07\,mag in $U$.)
During 1991-1998 22 measurements of the check star HD\,192987
were obtained. The mean values (N = 22) of the magnitude 
differences (HD\,192538 minus HD\,192987) and their standard
deviations are  $\Delta V = 0.002 \pm0.008$ mag
and $\Delta B = 0.088 \pm0.009$ mag. 
The scatter in Fig.\,1 is greater than the observational error
($\sim 5\,\sigma$\,in\,$B!$) and, therefore, probably contains real 
erratic variations with small amplitudes. 

In 1997, when the last 
peak in the IR was observed (Williams 1997), no
photometric effect can be seen, apart
from small-amplitude random variations. Their origin       
should arise in the continuum, as the plots in Fig. 2 suggest:
There are some correlations ($r = 0.58$ for $B$ and $V$
and $r = 0.64$ for $U$ and $B$) 
between the lightcurves in each of the three
passbands, which would be difficult to explain
by variability of emission lines. The origin  
of the small-amplitude random continuum 
variations of WR\,137 is possibly related to dynamical wind
instabilities, resulting in temperature effects  
at the ``pseudo-photospheric'' level.

\begin{table}
\caption[]{Differential photometry of WR\,140
(= HD 193793)  --  in the sense comparison star
HD 193888 minus WR\,140. Orbital phases are
calculated with $P = 2900$ d and $T_0 = 1985.26$.}
\renewcommand{\baselinestretch}{0.8}
\small
\begin{tabular}{rccccc}
\hline
  \noalign{\smallskip}
Year  &  JD-2400000 & orb.   &    $\Delta V$   &     $\Delta B$    &    $\Delta U$ \\
      &             & phase  &      [mag]      &       [mag]       &  [mag]      \\
  \noalign{\smallskip}
\hline
  \noalign{\smallskip}
1991  &  48540.340  &  0.821 &     1.685 &  1.266 &  1.377 \\
      &  48563.250  &  0.829 &     1.650 &  1.247 &  1.354 \\
      &  48565.256  &  0.829 &     1.653 &  1.248 &  1.349 \\
  \noalign{\smallskip}
\hline
  \noalign{\smallskip}
1992  &  48781.539  &  0.904 &     1.659 &  1.249 &  1.358 \\
      &  48860.432  &  0.931 &     1.672 &  1.255 &  1.359 \\
      &  48861.420  &  0.932 &     1.678 &  1.256 &  1.351 \\
      &  48862.457  &  0.932 &     1.680 &  1.268 &  1.349 \\
      &  48863.434  &  0.932 &     1.671 &  1.243 &  1.348 \\
      &  48865.387  &  0.933 &     1.686 &  1.267 &  1.349 \\
  \noalign{\smallskip}
\hline
  \noalign{\smallskip}
1993  &  49212.469  &  0.053 &     1.652 &  1.232 &  1.333 \\
      &  49220.422  &  0.055 &     1.647 &  1.238 &  1.331 \\
      &  49221.379  &  0.056 &     1.649 &  1.243 &  1.333 \\
      &  49222.410  &  0.056 &     1.654 &  1.241 &  1.337 \\
      &  49223.396  &  0.056 &     1.645 &  1.228 &  1.323 \\
      &  49224.413  &  0.057 &     1.655 &  1.238 &  1.330 \\
      &  49233.362  &  0.060 &     1.647 &  1.236 &  1.344 \\
      &  49234.373  &  0.060 &     1.640 &  1.232 &  1.332 \\
  \noalign{\smallskip}
\hline
  \noalign{\smallskip}
1994  &  49582.463  &  0.180 &     1.661 &  1.250 &  1.339 \\
      &  49584.433  &  0.181 &     1.669 &  1.252 &  1.347 \\
      &  49585.395  &  0.181 &     1.662 &  1.243 &  1.342 \\
      &  49586.402  &  0.182 &     1.679 &  1.259 &  1.345 \\
      &  49586.422  &  0.182 &     1.675 &  1.260 &  1.348 \\
      &  49587.385  &  0.182 &     1.665 &  1.246 &  1.343 \\
      &  49587.402  &  0.182 &     1.665 &  1.246 &  1.343 \\
      &  49589.392  &  0.183 &     1.670 &  1.252 &  1.343 \\
      &  49589.411  &  0.183 &     1.668 &  1.250 &  1.345 \\
      &  49594.392  &  0.184 &     1.664 &  1.250 &  1.342 \\
      &  49594.410  &  0.184 &     1.659 &  1.245 &  1.339 \\
      &  49594.428  &  0.184 &     1.665 &  1.247 &  1.340 \\
      &  49595.345  &  0.185 &     1.671 &  1.258 &  1.342 \\
      &  49596.352  &  0.185 &     1.670 &  1.246 &  1.329 \\
      &  49666.296  &  0.209 &     1.663 &  1.244 &  1.343 \\
  \noalign{\smallskip}
\hline
  \noalign{\smallskip}
1995  &  49922.534  &  0.297 &     1.656 &  1.228 &  1.324 \\
      &  49947.473  &  0.306 &     1.672 &  1.259 &  1.350 \\
      &  49949.451  &  0.307 &     1.669 &  1.258 &  1.348 \\
      &  49953.368  &  0.308 &     1.675 &  1.249 &  1.360 \\
      &  49954.396  &  0.308 &     1.671 &  1.260 &  1.359 \\
      &  49958.414  &  0.310 &     1.676 &  1.242 &  1.342 \\
      &  49976.408  &  0.316 &     1.673 &  1.249 &  1.353 \\
      &  49977.368  &  0.316 &     1.663 &  1.242 &  1.350 \\
      &  49998.263  &  0.324 &     1.661 &  1.243 &  1.355 \\
      &  50001.288  &  0.325 &     1.661 &  1.240 &  1.352 \\
      &  50003.256  &  0.325 &     1.665 &  1.244 &  1.357 \\
  \noalign{\smallskip}
\hline
  \noalign{\smallskip}
1996  &  50313.427  &  0.432 &     1.674 &  1.249 &  1.354 \\
      &  50317.392  &  0.433 &     1.675 &  1.250 &  1.362 \\
      &  50318.402  &  0.434 &     1.673 &  1.258 &  1.360 \\
      &  50321.401  &  0.435 &     1.677 &  1.259 &  1.354 \\
      &  50322.369  &  0.435 &     1.676 &  1.263 &  1.355 \\
      &  50359.295  &  0.448 &     1.668 &  1.248 &  1.361 \\
  \noalign{\smallskip}
\hline
  \noalign{\smallskip}
1997  &  50676.407  &  0.557 &     1.677 &  1.277 &  1.372 \\
      &  50711.345  &  0.569 &     1.674 &  1.277 &  1.387 \\
      &  50714.301  &  0.570 &     1.672 &  1.277 &  1.391 \\
      &  50730.294  &  0.576 &     1.664 &  1.265 &  1.368 \\
  \noalign{\smallskip}
\hline
  \noalign{\smallskip}
1998  &  51046.413  &  0.685 &     1.678 &  1.272 &  1.371 \\
      &  51049.369  &  0.686 &     1.689 &  1.278 &  1.381 \\
      &  51050.356  &  0.686 &     1.688 &  1.279 &  1.385 \\
      &  51053.389  &  0.687 &     1.689 &  1.275 &  1.377 \\
      &  51054.365  &  0.688 &     1.689 &  1.277 &  1.382 \\
      &  51059.329  &  0.689 &     1.683 &  1.280 &  1.387 \\
  \noalign{\smallskip}
\hline
\end{tabular}
\end{table}

\begin{figure}
  \centering
  \resizebox{08cm}{!}{\includegraphics{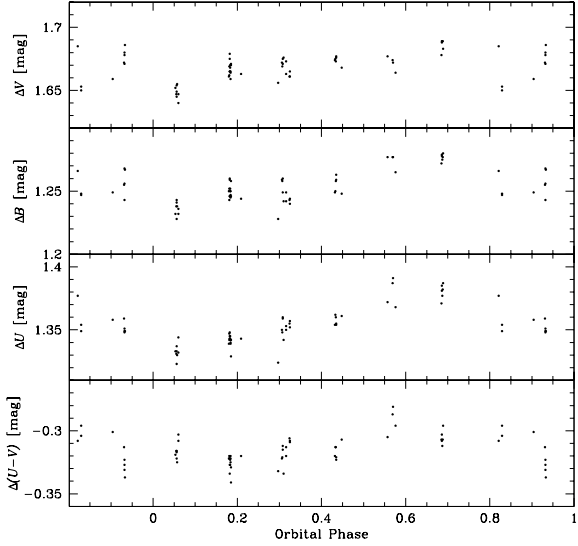}}
  \hfill
    \caption{Long-term light variations of WR\,140 (data from Table 3)}
\end{figure}

\begin{table}
\caption[]{Differential photometry of WR\,148
(= HD 197406)\\  --  in the sense comparison
star HD 197619 minus WR\,148.\\ Orbital phases
are calculated with $P = 4.317364$ d\\ and
$T_0 =$ JD 2432434.4 (Drissen et al. 1986).}
\renewcommand{\baselinestretch}{0.8}
\small
\begin{tabular}{rccccc}
\hline
  \noalign{\smallskip}
Year  &  JD-2400000 & orb.  &    $\Delta V$   &     $\Delta B$    &    $\Delta U$ \\
      &             & phase  &     [mag]      &        [mag]      &     [mag]     \\
  \noalign{\smallskip}
\hline
  \noalign{\smallskip}
1993   &  49212.502   &  0.19  &    $-$1.908 &  $-$2.313 &  $-$2.231 \\
       &  49220.454   &  0.03  &    $-$1.929 &  $-$2.331 &  $-$2.264 \\
       &  49221.402   &  0.25  &    $-$1.900 &  $-$2.287 &  $-$2.209 \\
       &  49222.432   &  0.49  &    $-$1.910 &  $-$2.302 &  $-$2.215 \\
       &  49223.419   &  0.72  &    $-$1.896 &  $-$2.288 &  $-$2.213 \\
       &  49224.434   &  0.95  &    $-$1.941 &  $-$2.334 &  $-$2.257 \\
       &  49233.384   &  0.03  &    $-$1.913 &  $-$2.309 &  $-$2.234 \\
       &  49234.395   &  0.26  &    $-$1.866 &  $-$2.253 &  $-$2.170 \\
       &  49235.373   &  0.49  &    $-$1.878 &  $-$2.288 &  $-$2.215 \\
  \noalign{\smallskip}
\hline
  \noalign{\smallskip}
1994   &  49582.507   &  0.89  &    $-$1.987 &  $-$2.382 &  $-$2.307 \\
       &  49584.468   &  0.35  &    $-$1.951 &  $-$2.344 &  $-$2.263 \\
       &  49585.424   &  0.57  &    $-$1.884 &  $-$2.222 &  $-$2.134 \\
       &  49586.454   &  0.81  &    $-$1.923 &  $-$2.326 &  $-$2.248 \\
       &  49587.429   &  0.03  &    $-$1.955 &  $-$2.348 &  $-$2.270 \\
       &  49589.442   &  0.50  &    $-$1.951 &  $-$2.343 &  $-$2.271 \\
       &  49594.476   &  0.67  &    $-$1.928 &  $-$2.329 &  $-$2.251 \\
       &  49595.373   &  0.87  &    $-$1.973 &  $-$2.360 &  $-$2.281 \\
       &  49596.375   &  0.10  &    $-$1.962 &  $-$2.353 &  $-$2.272 \\
  \noalign{\smallskip}
\hline
  \noalign{\smallskip}
1996   &  50313.505   &  0.21  &    $-$1.942 &  $-$2.337 &  $-$2.263 \\
       &  50317.435   &  0.12  &    $-$1.955 &  $-$2.365 &  $-$2.284 \\
       &  50318.436   &  0.35  &    $-$1.964 &  $-$2.346 &  $-$2.266 \\
       &  50321.462   &  0.05  &    $-$1.947 &  $-$2.321 &  $-$2.247 \\
       &  50322.404   &  0.27  &    $-$1.922 &  $-$2.307 &  $-$2.222 \\
       &  50358.324   &  0.59  &    $-$1.899 &  $-$2.292 &  $-$2.209 \\
  \noalign{\smallskip}
\hline
  \noalign{\smallskip}
1997   &  50676.447   &  0.27  &    $-$1.922 &  $-$2.284 &  $-$2.201 \\
       &  50714.327   &  0.05  &    $-$1.913 &  $-$2.281 &  $-$2.212 \\
       &  50730.322   &  0.75  &    $-$1.876 &  $-$2.250 &  $-$2.175 \\
       &  50731.317   &  0.98  &    $-$1.905 &  $-$2.272 &  $-$2.199 \\
       &  50732.292   &  0.21  &    $-$1.883 &  $-$2.249 &  $-$2.170 \\
  \noalign{\smallskip}
\hline
  \noalign{\smallskip}
1998   &  51046.452   &  0.98  &    $-$1.942 &  $-$2.316 &  $-$2.245 \\
       &  51049.406   &  0.66  &    $-$1.933 &  $-$2.303 &  $-$2.229 \\
       &  51050.390   &  0.89  &    $-$1.930 &  $-$2.302 &  $-$2.207 \\
       &  51050.405   &  0.89  &    $-$1.924 &  $-$2.304 &  $-$2.212 \\
       &  51053.419   &  0.59  &    $-$1.930 &  $-$2.299 &  $-$2.220 \\
       &  51054.392   &  0.81  &    $-$1.924 &  $-$2.288 &  $-$2.218 \\
       &  51058.419   &  0.75  &    $-$1.936 &  $-$2.303 &  $-$2.226 \\
       &  51059.399   &  0.97  &    $-$1.949 &  $-$2.321 &  $-$2.245 \\
  \noalign{\smallskip}
\hline
\end{tabular}
\end{table}

\subsection{WR\,140}

WR\,140 = HD 193793 (WC7 + O4-5) is another  
periodic dust maker.
Williams et al. (1978, 1987a, 1987b, 1990) and 
Williams (1997) reported variations in the IR,
revealing brightenings in 1977, 1985, and in 1993,
which they attributed to the building of dust 
grains in the WR\,140 wind with a period of
7.94 yr. The re-occurrence of the heated dust has been
interpreted as due to wind-wind interaction in a binary
system. Earlier spectroscopic studies failed to 
reveal the binary motion. However, a re-analysis 
of earlier published radial velocities and using the 
period in the IR (7.94 yr) led to a successful 
determination of the orbit (Williams et al. 1987c).
It was found that the grain formation coincides with 
the periastron passage (PP) in the system (actually
occurring before PP). This
discovery was later confirmed by Moffat et al. (1987)
and now presents the basic model for studies of
WR\,140. Williams et al. (1990) and van der Hucht et
al. (1991) reported on variability of WR\,140 at
X-ray, UV, IR and radio-wavelengths.
Our photometry of WR\,140 is presented in Table 3
and the light curves are shown in Fig. 3. From 
Fig. 3, there is clear evidence for a dip in
the light in 1993, between orbital phases $\sim$0.9
and 1.1. The dip is seen in all passbands and should therefore be
due to continuum light attenuation. The amplitude
of the ``eclipse'' in the $V$ passband is 0.03 mag.

\begin{figure}
  \centering
  \resizebox{8cm}{!}{\includegraphics{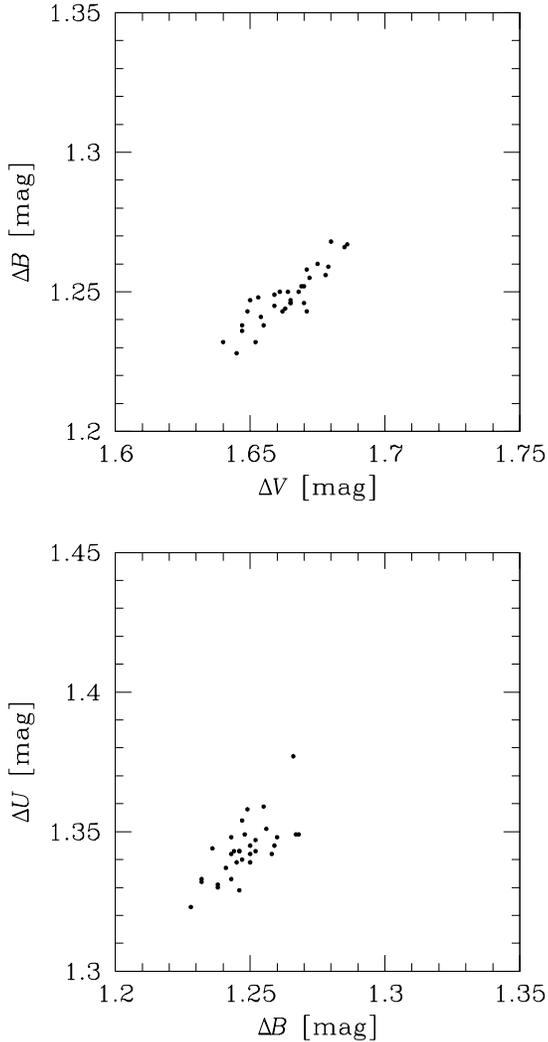}}
  \hfill
    \caption{WR\,140. Random light variability correlations for 1994
    (data from Table 3)}
\end{figure}

Two remarkable features are  to be mentioned. First, the very 
broad shape of the light minimum, assuming a smooth trend 
between yearly data.  After 1993, the light 
gradually increased to reach the ``pre-eclipse''
level in 1997, or even 1998. Considering the ``eclipse'' to be caused by 
an obscuration of the star(s) by the wind, the light curves
strongly suggest that a dust envelope was built up
around the WR star by the wind-wind interaction 
at the PP, which was gradually dispersed in
the following years. Possibly it is the same
dust observed in the IR when still heated.  
Second, it is apparent (Fig.~3) that the
amplitude of the eclipse increases towards
shorter wavelengths. In the lower panel of Fig.~3,  
the variation of the colour $\Delta(U - V)$ is shown,         
which is in the sense: WR\,140 colour gets redder
when its light is attenuated. This conclusion is  
easily obtained when considering the magnitudes of  
the comparison star HD 193888, which are: $V = 8.54$, 
$B-V = -0.07$, $U-B = -0.25$, and $U-V = -0.32$.
The amplitude of the colour variation in $U-V$ of WR\,140 is 
0.04 mag (again Fig.~3).        
As is well known, the main source of opacity, (non-relativistic) 
electron scattering, has no effect on colours.    
Thus, in this case electron scattering alone
is not sufficient, and an additional opacity source should
be introduced, possibly Rayleigh (or Mie) scattering by small 
carbon dust particles.

Occasional ``eclipses'' have been observed in the carbon-rich
late-type WC stars WR\,103, WR\,113, and WR\,121
(for a history of ``eclipses'' see Veen et al. 1998).
In these cases the ``eclipses'' were caused by occasional
formation of dust in the line-of-sight. Although dust
formation in the winds of late-type WC stars is now
well established, the problem with grain condensation in the
very hostile environments where the grains are believed to form
remains unsolved. Clearly, a trigger is needed to start 
the grain formation. In the case of WR\,140, this could be
the shock compression in the colliding winds at PP.
We assume that the fading of WR\,140 shortly after PP is due to
dust condensation in the wind of the WC star. 
After the condensation ceases the star brightens again because
the dust is blown away and gradually dispersed.
The ``eclipses'' studied by Veen et al. (1998) have typical
amplitudes of several tenths of a magnitude and last from
several days up to a month. In contrast, the amplitude of the
light dip in WR\,140 is much smaller and the recovery of
brightness lasts several years. This implies continuing
supply (expanding from the PP production + new?)
of dust, even 2 -- 3 years after PP. If there is new dust,
this would be really surprising, since the trigger seems
no longer to be effective. Following the procedure of 
Veen et al. (1998, using their equations (5), (6), and (7))
and taking the terminal velocity $v_\infty = 
2900\,\mathrm{km\,s}^{-1}$ from Eenens \& Williams (1994),
we obtain for the distance $R_\mathrm{cc}$ of the dust
formation region from the WC star in WR\,140:
$R_\mathrm{cc} \sim 300\,000\,R_{\sun}$. This is only
a rough estimate, but it is much larger than the
respective distances for all ``eclipses'' studied by
Veen and co-workers. It is also much larger than the radius
of the shell of WR\,140 obtained by Williams et al. (1987a)
which is $R_{shell} = 1490\,R_{\sun}$. Taking for the
carbon particle density $1.85\,\mathrm{g\,cm}^{-3}$
we get for the dust mass production rate (over unit area)
the value $\dot{M}_{d} = 2\,10^{-13}\, 
\mathrm{kg\,m}^{-2}\,\mathrm{s}^{-1}$. These results should
be taken with caution because of the small amplitude of
the ``eclipse'' in WR\,140 and of possible deviations from
the model used (e.g. continued supply of dust after PP).

WR\,140 was observed photometrically during PP in 1977 by
Fernie (1978) but no changes of brightness were found.
This is likely due to his low precision data.

Like WR\,137, the observations of WR\,140 also
show small-amplitude, day-to-day random light
variations (amplitudes up to 0.02 mag), in addition
to the eclipse variation. Fig.~4 shows the correlations 
of the random light variations in $UBV$, indicating 
that they are likely due to continuum rather than emission 
line variations (similar to WR 137, Fig. 2). Dynamical
wind instabilities could be the origin, as in WR\,137.
Moffat \& Shara (1986) suggested a 6.25 d period 
for the light variations they observed in WR\,140, 
which, however, does not fit our data. 
Our observations during 1991 - 1998 cover 90\%
of the orbit. It remains to be seen whether
the forthcoming PP in 2001 will repeat the light
curve so far observed.

\begin{figure*}
  \centering
  \resizebox{17cm}{!}{\includegraphics{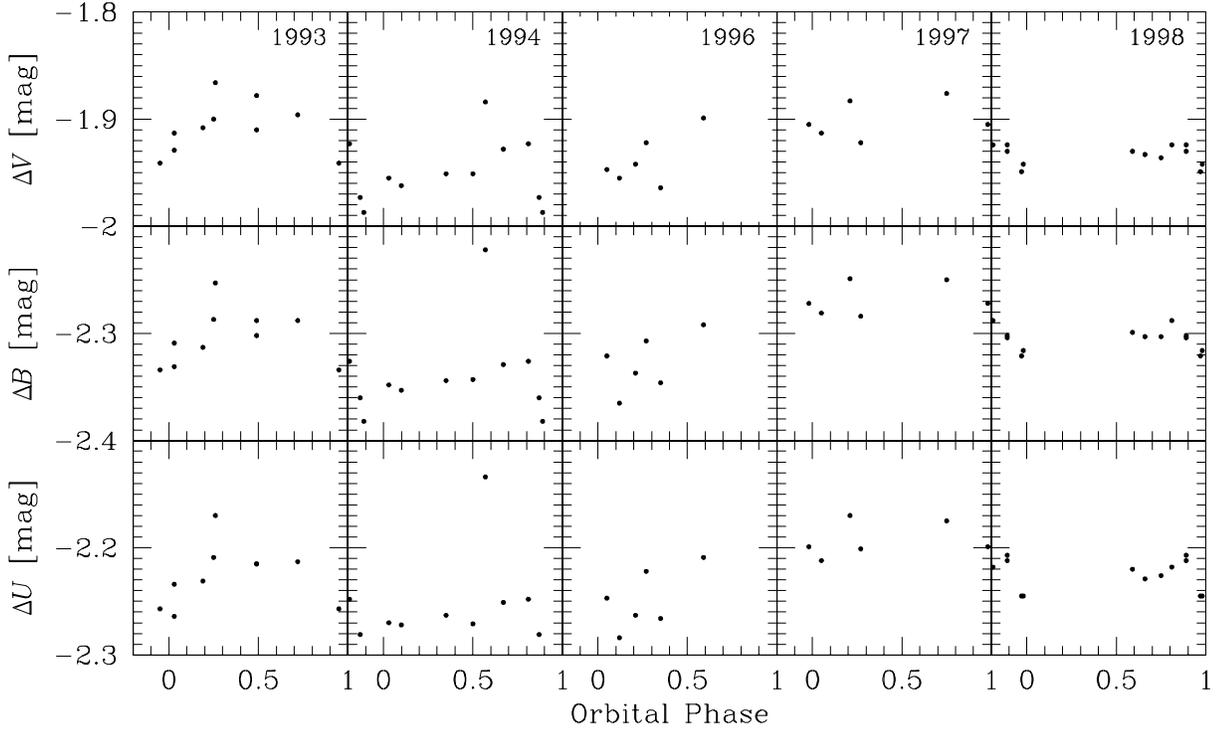}}
  \hfill
    \caption{Light curves of WR\,148
     (data from Table 4)}
\end{figure*}

\subsection{WR\,148}

\begin{table*}
\caption[]{ Photometry of WR\,153 (= HD 211853).
The comparison star is HD 211430\\ with $V = 7.465$,
$B - V = -0.054$, and $U - B = -0.490$.\\ Orbital
phases "1" are calculated with $P1 = 6.6884$ d
and $T_0 =$ JD 2443690.32 (Massey 1981),\\ orbital
phases "2" with $P2 = 3.4696$ d and
$T_0 =$ JD 2443689.16 (Annuk 1994)}
   \setlength\tabcolsep{12pt}
\renewcommand{\baselinestretch}{0.8}
\small
\begin{tabular}{rcccccccc}
\hline \noalign{\smallskip}
Year  &  JD-2400000 & \multicolumn{2}{c}{Orb. Phase}  &    $V$   &     $B$    &    $U$  & $B-V$ & $U-B$ \\
  &    &    "1"         & "2"   &  [mag]  &    [mag]   &  [mag] &    [mag]   &  [mag]   \\
  \noalign{\smallskip}
\hline
  \noalign{\smallskip}
1991   &  48448.527  & 0.41 & 0.73  & 8.979 & 9.352 & 8.749 & 0.373  & $-$0.603 \\
       &  48510.472  & 0.67 & 0.59  & 9.001 & 9.373 & 8.765 & 0.372  & $-$0.608 \\
       &  48510.487  & 0.68 & 0.59  & 8.996 & 9.368 & 8.766 & 0.372  & $-$0.602 \\
       &  48510.504  & 0.68 & 0.60  & 8.995 & 9.372 & 8.769 & 0.377  & $-$0.603 \\
       &  48510.518  & 0.68 & 0.60  & 8.992 & 9.370 & 8.774 & 0.378  & $-$0.596 \\
       &  48511.524  & 0.83 & 0.89  & 8.992 & 9.372 & 8.771 & 0.380  & $-$0.601 \\
       &  48511.538  & 0.83 & 0.89  & 8.995 & 9.375 & 8.775 & 0.380  & $-$0.600 \\
       &  48511.551  & 0.83 & 0.90  & 8.995 & 9.372 & 8.777 & 0.377  & $-$0.595 \\
       &  48511.567  & 0.84 & 0.90  & 8.994 & 9.377 & 8.775 & 0.383  & $-$0.602 \\
       &  48511.580  & 0.84 & 0.91  & 8.996 & 9.379 & 8.784 & 0.383  & $-$0.595 \\
       &  48511.591  & 0.84 & 0.91  & 9.000 & 9.380 & 8.788 & 0.380  & $-$0.592 \\
       &  48512.519  & 0.98 & 0.18  & 9.042 & 9.428 & 8.818 & 0.386  & $-$0.610 \\
       &  48513.479  & 0.12 & 0.45  & 9.052 & 9.445 & 8.855 & 0.393  & $-$0.590 \\
       &  48514.502  & 0.28 & 0.75  & 8.978 & 9.353 & 8.737 & 0.375  & $-$0.616 \\
       &  48538.446  & 0.86 & 0.65  & 8.972 & 9.363 & 8.752 & 0.391  & $-$0.611 \\
       &  48539.464  & 0.01 & 0.94  & 9.068 & 9.462 & 8.862 & 0.394  & $-$0.600 \\
  \noalign{\smallskip}
\hline
  \noalign{\smallskip}
1993   &  49220.514  & 0.83 & 0.23  & 8.979 & 9.367 & 8.751 & 0.388  & $-$0.616 \\
       &  49221.490  & 0.98 & 0.52  & 9.086 & 9.478 & 8.875 & 0.392  & $-$0.603 \\
       &  49222.523  & 0.13 & 0.81  & 8.972 & 9.363 & 8.743 & 0.391  & $-$0.620 \\
       &  49223.488  & 0.28 & 0.09  & 8.991 & 9.380 & 8.768 & 0.389  & $-$0.612 \\
       &  49224.519  & 0.43 & 0.39  & 8.972 & 9.361 & 8.750 & 0.389  & $-$0.611 \\
       &  49233.431  & 0.76 & 0.96  & 9.028 & 9.421 & 8.805 & 0.393  & $-$0.616 \\
       &  49233.446  & 0.77 & 0.96  & 9.929 & 9.418 & 8.808 & 0.389  & $-$0.610 \\
       &  49234.449  & 0.92 & 0.25  & 8.990 & 9.383 & 8.766 & 0.393  & $-$0.617 \\
       &  49234.462  & 0.92 & 0.25  & 8.986 & 9.382 & 8.761 & 0.396  & $-$0.621 \\
       &  49234.474  & 0.92 & 0.25  & 8.992 & 9.379 & 8.760 & 0.387  & $-$0.619 \\
       &  49234.490  & 0.92 & 0.26  & 8.994 & 9.382 & 8.764 & 0.388  & $-$0.618 \\
       &  49234.503  & 0.93 & 0.26  & 8.994 & 9.380 & 8.760 & 0.386  & $-$0.620 \\
       &  49234.517  & 0.93 & 0.27  & 8.989 & 9.382 & 8.764 & 0.393  & $-$0.618 \\
       &  49235.416  & 0.06 & 0.53  & 9.072 & 9.470 & 8.858 & 0.398  & $-$0.612 \\
       &  49235.429  & 0.06 & 0.53  & 9.076 & 9.478 & 8.862 & 0.402  & $-$0.616 \\
       &  49235.441  & 0.07 & 0.54  & 9.085 & 9.483 & 8.858 & 0.398  & $-$0.625 \\
  \noalign{\smallskip}
\hline
  \noalign{\smallskip}
1994   &  49582.539  & 0.96 & 0.58  & 9.040 & 9.431 & 8.835 & 0.391  & $-$0.596 \\
       &  49584.494  & 0.25 & 0.14  & 8.993 & 9.381 & 8.770 & 0.388  & $-$0.611 \\
       &  49585.476  & 0.40 & 0.42  & 8.994 & 9.380 & 8.771 & 0.386  & $-$0.609 \\
       &  49586.502  & 0.55 & 0.72  & 8.976 & 9.368 & 8.763 & 0.392  & $-$0.605 \\
       &  49587.477  & 0.70 & 0.00  & 9.009 & 9.393 & 8.787 & 0.384  & $-$0.606 \\
       &  49589.498  & 0.00 & 0.58  & 9.051 & 9.448 & 8.844 & 0.397  & $-$0.604 \\
       &  49594.505  & 0.75 & 0.02  & 9.022 & 9.418 & 8.817 & 0.396  & $-$0.601 \\
       &  49595.417  & 0.89 & 0.29  & 8.985 & 9.375 & 8.763 & 0.390  & $-$0.612 \\
       &  49596.402  & 0.03 & 0.57  & 9.069 & 9.464 & 8.845 & 0.395  & $-$0.619 \\
       &  49666.396  & 0.50 & 0.74  & 8.984 & 9.379 & 8.767 & 0.395  & $-$0.612 \\
  \noalign{\smallskip}
\hline
  \noalign{\smallskip}
1995   &  49975.516  & 0.72 & 0.84  & 9.002 & 9.391 & 8.781 & 0.389  & -0.610 \\
       &  49977.524  & 0.02 & 0.42  & 9.054 & 9.436 & 8.832 & 0.382  & -0.604 \\
  \noalign{\smallskip}
\hline
\end{tabular}
\end{table*}

WR\,148 (= HD 197406, WN8 + c?) is a single-line     
spectroscopic binary, possibly hosting a compact companion.
The star has been studied by Bracher (1979). She 
determined the orbital period as $P = 4.3174$ d  
and also found light variations with the same    
period and an amplitude of 0.04 mag in $V$. Further   
spectroscopic studies by Moffat \& Seggewiss   
(1979, 1980) revealed an unusually low mass  
function of the system, which was later  
confirmed by Drissen et al. (1986): f(m) = 0.28 M$_{\sun}$. 
WR\,148 has also an exceptionally large 
distance from the galactic plane, $z = 500 - 800$ pc  
(Moffat \& Isserstedt 1980; Dubner et al. 1990).
Smith et al. (1996) found that WR\,148 is a WN8 star.  
The low mass function and high $z$ value led
Moffat \& Seggewiss (1980) to advance the idea  
that WR\,148 harbours a compact companion as product
of a supernovae explosion some 5 Myr  
ago. In their model, the companion is orbiting
within the WR envelope. As the companion orbits around the WR star
the projected envelope density varies. This is the origin of the
 light variations of WR\,148, because electron scattering occurs in this 
envelope. 
Photometric studies by Antokhin (1984),
Moffat \& Shara (1986), and Marchenko et al.  
(1996) confirmed the light variations found by Bracher (1979) with an
amplitude of 0.03 mag in $V$ and also point
to the very ``noisy'' appearence of
the light curve. (With the ephemeris 
of Drissen et al. (1986), the light minimum 
occurs at phase zero with the WR star in
front). Marchenko et al. (1996) noted the
unusual wide-shaped light minimum, quite 
different from other known WR + O systems 
with atmospheric eclipses and a V-shaped light
minimum (Lamontagne et al. 1996). For WR\,148,
Marchenko et al. suggested that the secondary 
light arises from an extended hot cavity 
in the WR envelope, near the companion, 
and which is ionized by X-rays. According 
to Marchenko et al., the rather weak
X-ray source observed in WR\,148
(Pollock et al. 1995)
may be explained by the hot X-ray 
cavity being locally embedded in the WR envelope. 
Presently, the evolutionary status of
WR\,148 remains unclear and the companion
could be either a B2-B4 III-V star or a 
relativistic object (as deduced from the mass function,
Marchenko et al. 1996).  
                                                     
Our photometry is presented in Table 4 and
the light curves are shown in Fig.~5,
plotted with the ephemeris of Drissen et al.
(1986). From Fig.~5 it is apparent that our
light curves in 1993 are similar to the 
light curves published by Moffat \& Shara (1986).
The minimum occurs at phase zero. The 
1994 light curves, however, show a remarkable
change in their shape and mean light level.
Random light variations, already noted in 
other works, could well contribute to the 
disturbance of the light curve shape, but
it is unlikely that they would change the
mean light. Furthermore, long-term changes 
in mean light appear to be 
correlated in $U$, $B$, and $V$ (Fig.~5). Therefore,
they too should be due to changes in 
continuum light. 

There is a strong
evidence for a long-term variation of the
mean light. Although the time-base is too short,               
there are some indications that the long-term variation  
is periodic, possibly with a cycle of about 4 years. 
Marchenko et al. (1998b) point to a possible
``overall brightening'' of WR\,148 in 1994 and 
1995. As shown in Fig.\,5, it is obvious that
in 1993 the mean light was even some 0.05 mag higher,
as in 1994. This long-term variation completely masks 
the short-term binary variations if the whole data set 
is depicted in one plot. Therefore we
plotted the data separately for each year in Fig.~5.
 
Taking into account the model of Marchenko et al.(1996), 
the long-term light variations in WR\,148 
could be due to variations of the size 
of the hot X-ray cavity.
Further conclusions at that time seem premature.  
A comment should be given on the observation at
JD 2449585.4, phase = 0.57 (the companion $\sim$ in front),
which strongly deviates from the regular light curve of 1994.       
As we can exclude observational errors as a reason for 
this measurement, it has to have some astrophysical origin.
For instance, an event
of accretion onto a compact companion could be 
invoked to explain this flare-like burst. 

Flickering and flaring of
WR\,148 on different time-scales have been
reported by Antokhin \& Cherepashchuck (1989),
Zhilyaev et al. (1995) and
Khalack \& Zhilyaev (1995). Matthews et al.
(1992) looked for flares in the WR star EZ Canis
Majoris (WR\,6 = HD 50896, WN5) and reported one
flare event. Flare-type activity of EZ CMa was also
observed by Duijsens et al. (1996). This star is in many
respects similar to WR\,148, e.g. showing
light variations with a 3.77 d period,
long-term changes in the light curve, and a
possible WR + c binary status (Firmani et al. 1980;
Balona et al. 1989; Duijsens et al. 1996).    

\begin{figure*}  
  \centering    
  \resizebox{12cm}{!}{\includegraphics{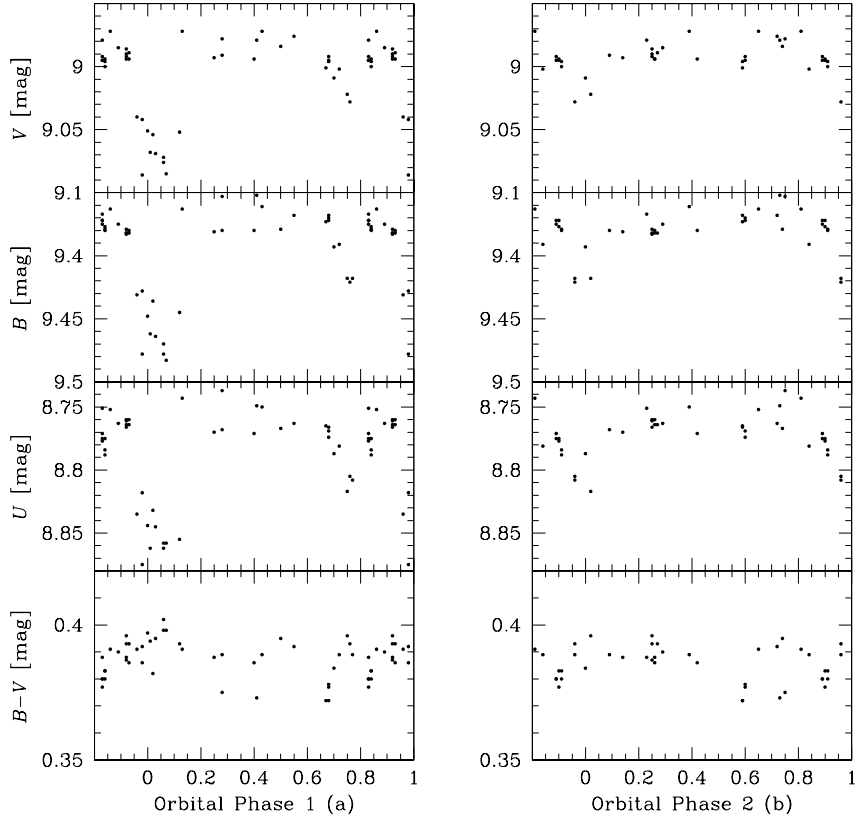}}  
  \hfill  
    \caption{Light curves of WR\,153 with (a) $P1 = 6.6884$ d  
    and (b) $P2 = 3.4696$ d (data from Table~5)}
\end{figure*}
                                                     
\subsection{WR\,153}
WR\,153 (= HD 211853 = GP Cep) is a quadruple system (Massey 1981) 
with orbital periods 6.6884 d (pair A, WR + O) and
3.4696 d (pair B, WR + O or O + O). Earlier spectroscopic
studies by Hiltner (1945) and Bracher (1968)
revealed radial velocity variations due to
binary motion with a period of 6.68 d. Panov \&
Seggewiss (1990) reanalysed Hiltner`s velocity
data and found evidence for two WR stars, one
in each pair. WR\,153 has been observed photometrically
by Hjellming \& Hiltner (1963), Stepien (1970),
Moffat \& Shara (1986), Panov \& Seggewiss (1990),
and Annuk (1994), all detecting eclipses
with both periods, 6.6884 d and 3.47 d.
Finally, Annuk (1994) refined the second period
to 3.4696 d, in agreement with the velocity
data of Massey (1981). However, in the recent 
analysis of WR star light curves by Lamontagne et al. 
(1996) the 3.47 d variation of pair B could not
unambiguously be extracted from their data. 

Our photometry of WR\,153 is presented in
Table 5 and the light curves are shown in
Fig.~6a and Fig.~6b, with the 6.6884 d and 3.4696 d
periods, respectively. From Fig.~6, our
data are consistent with the ephemeris of 
Massey (1981) and Annuk (1994), respectively.
Since the true shape of both 
light curves is unknown, no allowance is made for the
3.47 d period in Fig.~6a, where it is
superimposed on the 6.69 d light variations.
In Fig.~6b, the data points around the
6.69 d period minimum (at phases from 0.96 to 0.13 in 
Fig.~6a) have been removed. 

The light curve with
the 6.69 d period (pair A) is probably due to
an atmospheric eclipse (only one, V-shaped
light minimum!). In pair B, two light minima
are seen due to a core eclipse in that pair.
Independent evidence can be obtained from 
the HIPPARCOS photometry data. We made a period
search, using 122 data-points from HIPPARCOS. The   
analysis was performed with the PERIBM 
procedure, developed at the Astronomical
Institute of the University of Vienna (latest version from:\\
ftp://dsn.astro.univie.ac.at./pub/PERIOD98/current/).
Our analysis clearly shows that there are peaks at
$f1 = 0.5763641$ d$^{-1}$, corresponding to a
1.735 d period, and at $f2 = 0.1495867$ d$^{-1}$,
corresponding to the 6.69 d period. The 
1.735 d period is exactly 1/2 of the 
3.47 d period and the reason that it shows up 
in the amplitude spectrum is because of the
double-wave light curve (two eclipses in pair B),
consistent with our ground-based photometry.
Moffat \& Shara (1986) also deduced that pair A had a
single minimum at phase 0.00 ($P = 6.69$\,d) and pair B
had a double minimum at phases 0.00 and 0.50
($P = 3.47$ d).

\section{Conclusions}

Our photometry of WR\,137 reveals
only small amplitude ($\le$ 0.03 mag in the
$V$ passband) random light variations. No
periodicity could be found. These variations 
should be attributed to the continuum and they are
probably due to dynamical wind instabilities.  

WR\,140 exhibited remarkable light variations
and a shallow light dip is seen in
all passbands shortly after periastron
passage in 1993.  The light
attenuation lasted until 1997 or even 1998, 
probably because of a dust envelope built
around the WR star by wind-wind interaction
at periastron passage. The dust envelope was
gradually dispersed. From the wavelength
dependence of the light attenuation, we
find strong evidence for Rayleigh (Mie-like)
scattering, contributing to the opacity,
in addition to electron scattering.  

For WR\,148, our photometric study confirms   
the 4.317364 d light variation, but  
reveals occasional scatter, disturbing the  
light curve shape. On one occasion, we see a
flare-like event at a phase when the companion
is in front. Our photometry reveals long-term
variations of the mean light and, possibly, of
the amplitude of the regular variation.        
There is some evidence for a periodicity of
the long-term light variation and a 4 year
cycle cannot be ruled out.    

Our photometry of WR\,153 is consistent  
with the quadruple model for this star and
both the 6.6884 d and the 3.4696 d periods are seen      
in the light. In pair A (6.6884\,d period) we found      
evidence for an atmospheric eclipse, in
agreement with the results of other works, while in  
pair B (3.4696 d period) the eclipse
is probably photospheric (core eclipse).

\acknowledgements{
This research project was supported by the Deutsche 
Forschungsgemeinschaft DFG,\\ grant 436\,BUL 113/88/0.
In addition, M. Altmann is grateful to the DFG for grant Bo 779/21.
K.P. Panov kindly acknowledges the support by the Bulgarian National 
Science Foundation, grant F\,826. Thanks are due to J.S.W. Stegert 
for his participation in the observations. The authors are thankful
to the referee Dr. A.F.J. Moffat for his fruitful suggestions.
For our research we made with pleasure use of the SIMBAD data base 
in Strasbourg.}

{}
\end{document}